# Mutual friction in superfluid $^4$He near the $\lambda$-Line

Kerry Kuehn and Guenter Ahlers

*Department of Physics and Quantum Institute,*

*University of California, Santa Barbara, CA 93106, USA*

(November 12, 2001)

## Abstract

We present experimental results for the thermal resistivity $\rho$ of superfluid $^4$He along several isobars between saturated vapor pressure and the melting pressure. The measurements are for the temperature range $1 - T_c(q)/T_\lambda < t < 2 \times 10^{-5}$ and the heat-flux range $3 < q < 70$ $\mu$W/cm$^2$. Here $t \equiv 1 - T/T_\lambda$, $T_\lambda$ is the transition temperature in the limit of zero $q$, and $T_c$ is the transition temperature at finite $q$. The data suggest that the resistivity has an incipient singularity at $T_\lambda$ which can be described by the power law $\rho = (t/t_0)^{-(m\nu+\alpha)}$ where $t_0 = (q/q_0)^x$. However, the singularity is supplanted by the transition to a more highly dissipative phase at $T_c(q) < T_\lambda$. The results suggest a mild dependence of $m\nu + \alpha$ on $P$, but can be described quite well by $m\nu + \alpha = 2.76$, $x = 0.89$, and $q_0 = q_{0,0} - q_{0,1}P$ with $q_{0,0} = 401$ Wcm$^{-2}$ and $q_{0,1} = -5.0$ W cm$^{-2}$ bars$^{-1}$. The results imply that the Gorter-Mellink mutual friction exponent $m$ has a value close to 3.46 and is distinctly larger than the classical value $m = 3$. We suggest that the reason for this may be found in the nature of the counterflow close to $T_\lambda$, which is expected to involve turbulent normalfluid flow.

# I. INTRODUCTION

In an earlier paper [1] results of high-resolution measurements at saturated vapor pressure (SVP) of the thermal resistivity $\rho$ of a bulk sample of superfluid $^4$He below but close to the superfluid transition at $T_c$ were reported. Two interesting points were noted. First, the data suggested that $\rho$ has an incipient singularity at the zero-current $T_\lambda$ and *not* at the $q$-dependent $T_c$. This singularity is pre-empted by the transition to the normal phase at $T_c(q) < T_\lambda$ where $\rho$ is finite. Second, the exponent $m$ describing the $q$ dependence of $\rho$ differed markedly from the value $m = 3$ which was originally suggested by Gorter and Mellink [2] and which was later justified by Vinen [3] on the basis of simple dimensional analysis.

In the present paper we report results of new measurements which extend the previous results to several pressures between SVP and the melting curve. They confirm the earlier qualitative findings at SVP mentioned above. Over the entire pressure range the results can be described by a simple powerlaw for $\rho$ with significant pressure dependence only in the amplitude.

In the next section we describe the experimental apparatus. In Sect. III, the experimental measurement procedure and sources of potential systematic errors are discussed. Results for $\rho$ and their analysis are presenteed in Sect. IV. A summary and discussion is given in in section V.

# II. EXPERIMENTAL APPARATUS

The cryostat was a modified version of one used previously in our group [1]. A schematic diagram is shown in Fig. 1. We describe here only the features important to the present experiment, namely the refrigeration, the cell design and thermometry, and the equipment used for sample pressure measurement and control.

### A. Refrigeration

Cooling power was provided by a continuous $^4$He evaporator [4] which operated near 1.4 K. The thermal links between the cell mounting stage and the refrigeration stage were chosen so as to provided cooling power to the cell of 80$\mu$W at SVP (where $T_\lambda$ is near 2.18 K) and 40$\mu$W at 29 bar (where $T_\lambda$ is near 1.76 K).

### B. Cell design

The sample cell (denoted as cell III) was similar to the one used in the earlier work [1] (cell II). It was used during two cooldowns [5]. All of the merasurements at elevated pressure were made during the second cooldown. Results at SVP were obtained during both thermal cycles. A schematic diagram of the cell is given in Fig. 2. The cell consisted of two OFHC copper endplates separated by 0.485 cm by a stainless steel tube with 0.012 cm thick walls. The tube was silver soldered to the bottom endplate, leaving no gap between the sidewall and the protruding copper anvil; at the top endplate an indium o-ring seal was used near the bolt holes which left a tiny gap between the sidewall and the copper anvil. Both interior copper surfaces had an area of 1.267 cm$^2$ and were polished to an r.m.s. roughness of fifty nanometers or less.

### C. Cell thermometry

Magnetic susceptibiltity thermometers (MSTs) [6] were mounted on the cell top and bottom and also on two thin copper sideplanes which were used to probe the local fluid temperature at intermediate positions along the cell height. Germanium resistance thermometers, calibrated against a standard thermometer [7], were mounted on both the cell top and bottom for calibration of the MSTs. The temperature during a typical experimental run could be regulated to better than 10$^{-9}$K at SVP and roughly 5 × 10$^{-9}$K at elevated pressures. At elevated pressures it was more difficult to control the temperature, presumably due to adiabatic temperature changes of the sample fluid during pressure regulation.

### D. Sample preparation

All experiments were performed on a bulk cylindrical sample of ultra-pure $^4$He (0.5ppb $^3$He) of cross sectional area 1.267 cm$^2$ and height 0.485 cm. The helium was brought to the cell via a 0.025 cm inner-diameter stainless-steel capillary tube. A normally-closed valve mounted on the first isolation stage of the cryostat was used to isolate the fluid in the cell. To fill the cell, first the cryogenic valve actuator was pressurized to nearly 17 bar, opening the valve. Next, $^4$He gas at the desired pressure was introduced into the cell fill line. When the cell was filled with liquid of the desired pressure, the valve actuator was depressurized, closing the valve. Typically, the valve was kept above 4.2K until the actuator line was fully evacuated, after which the valve temperature was dropped to the isolation stage temperature.

For work at elevated pressures a hot volume was used to regulate the fluid pressure [8] after closing the cryogenic valve. An ac bridge method [9] was used to measure the pressure. Two arms of the bridge were provided by a Straty-Adams [10] pressure transducer and a reference capacitor mounted at low temperature on the second isolation stage (see Fig. 1). Two additional arms were provided by a ratio transformer at room temperature. The out-of-balance signal, measured by a lockin amplifier, was the input of a temperature controller which drove a 5k$\Omega$ metal-film heater mounted on the hot volume. In this manner the pressure could typically be controlled to better than half a microbar (see Sect. II E).

For SVP experiments, the cell was filled until a liquid-vapor interface was established in a copper reservoir mounted on the radiation shield (see Fig. 1). This shield was regulated at a temperature about 20 milliKelvin above $T_\lambda$ with slightly better than one microkelvin resolution. Consequently the interface pressure was stable to about $10^{-7}$ bars and $T_\lambda$ of the fluid in the cell was stable to about $10^{-9}$K.

### E. Pressure regulation

Time series of the pressure noise, measured both at SVP and at 28.8 bars, are shown in Fig. 3. At SVP the pressure was constant by virtue of being equal to the vapor pressure at the constant shield temperature. At the elevated pressure, the hot volume was used to regulate the pressure. The Nyquist frequency was 1/2 Hz. Also shown are histograms of the pressure-fluctuation probability-distributions. The SVP data have a wider distribution; the gaussian fit has a variance of $9 \times 10^{-14}$ bar$^2$. The 28.8 bar data have a narrower distribution; the gaussian fit has a variance of $7 \times 10^{-15}$ bar$^2$, corresponding to a root-mean-square pressure fluctuation of slightly less than 0.1 $\mu$bar. Also shown is the power spectral density (PSD) of each time series.

The pressure stability and noise were more than adequate to maintain the system on the desired isobar. More detrimental were the temperature fluctuations induced by adiabatic compression of the sample due to pressure fluctuations. One has $(\partial T/\partial P)_S = TV\alpha_P/C_P$ where $\alpha_P$ is the isobaric expansion coefficient and $C_P$ is the isobaric specific heat. Using the values $C_p \simeq 60$ J/mole K, $\alpha_P = -0.3$ K$^{-1}$, and $V \simeq 20$ cm$^3$/mole which are "typical" for the higher pressures, one has $(\partial T/\partial P)_S \simeq -0.02$ K/bar. This implies that a $10^{-7}$ bar pressure fluctuation will induce a temperature fluctuation of order 2 nK. Whereas this is comparable to the temperature noise of our thermometers, it is clear that any occasional larger pressure perturbations, which occurred due to external perturbations of unknown origin or due to deliberate changes in experimental setpoints, were detrimental to the measurements on isobars. Of course the SVP measurements were not affected by this problem.

### III. EXPERIMENTAL PROCEDURE

An experimental run consisted of applying a constant heat flux $q$ to the fluid while ramping its temperature upwards through the superfluid–normalfluid transition. The following procedure, an example of which is shown in Fig. 4a, was used to obtain data at each $q$ and for each pressure $P$.

The cell top temperature was regulated at a temperature a couple of microkelvin below the transition (A). Next, a heat flux of 15.8 nW/cm$^2$ was applied to the cell from the bottom (B). Several minutes were allowed for a steady state temperature configuration to develop. The cell top temperature was then ramped at a constant rate of 3 nK/second up and down through the transition (C). The heat flux was then turned off (D). This procedure served to determine $T_\lambda$, the transition temperature in the limit of very low heat flux.

The cell top temperature was then dropped to several microkelvin below the transition (E). After equilibration, a known heat flux, in this case 2.37 $\mu$W/cm$^2$, was applied to the cell bottom. The fluid was again allowed to reach its steady state temperature configuration (F). The cell top temperature was then ramped at a rate of 0.7 nK/second upwards through the transition (G).

A number of issues arose during a typical run which required special attention. Since we were interested in measuring very small differences in temperature, even a small error in the thermometer calibration would have obscured the measurements. Consequently, the temperature readings of the different thermometers had to be carefully aligned deep in the superfluid phase at two different temperatures, for example at sections (D) and (E), where under zero heat flux there should be no thermal gradient. Typically, this relative sensitivity calibration between the different thermometers was done before each ramp.

At SVP, the thermometers were quite stable. This was not the case at high pressures where the thermometers typically suffered more flux jumps and drift relative to one another. This is possibly due to a combination of imperfect magnetic shielding and increased thermometer sensitivity while operating closer to the Curie point of the paramagnetic salt pill. Small but constant relative drifts and flux jumps required corrections during data analysis and offered a source of systematic errors, particularly in the runs at high $P$ and very low $q$.

## IV. ANALYSIS AND RESULTS

The effect to be studied, namely the thermal resistivity in the superfluid phase in the presence of a heat flux, is clearly visible in Fig. 4b. There the upper and lower traces represent the temperatures at the lower and upper sideplanes respectively. The vertical line represents the time at which the fluid at the cell bottom reached $T_c$. The small temperature difference $\delta T$ between the sideplanes to the left of this line is due to mutual friction and is the subject of our studies. Slightly to the right of this line, the interface between the superfluid and normal phases (i.e. the local $T_c(q)$) passed the bottom sideplane. The resistivity of the fluid can be determined from the sideplane spacing $L$, the heat flux $q$, and the temperature difference $\delta T \equiv T_{st} - T_{sb}$ between the top sideplane temperature and the bottom sideplane temperature via the formula

$$\rho \equiv -\frac{\delta T}{Lq} \quad (1)$$

This assumes a linear temperature profile in the fluid which is an excellent approximation for the small temperature differences which are involved.

Figure 5a shows $\rho$ at SVP for several values $q$ as a function of the reduced temperature $t \equiv 1 - \bar{T}/\bar{T}_\lambda$. Here $\bar{T}$ is the average of the two sideplane temperatures and $T_\lambda$ is the zero-$q$ transition temperature on the isobar in question and at the vertical position halfway between the two sideplanes. Inspection of the figure suggests that the results agree well with previous measurements [1] in that they exhibit a power law dependence on $t$ with a $q$-independent exponent (the slope of lines through the data sets) and a strongly $q$-dependent amplitude. Figure 5b shows $\rho$ at constant $q$ for several values of $P$. The data suggest that $\rho$ is given by a powerlaw with a $P$-independent exponent and an amplitude which increases with $P$. It should be emphasized that from both plots it is apparent that $\rho$ has an incipient singularity at $T_\lambda$, the transition temperature in the $q \to 0$ limit, and not at $T_c(q)$ where it remains finite.

To interpret these results quantitatively, we use as a guide the formula suggested by

Gorter and Mellink [2,11] to describe the thermal gradient arising as a consequence of mutual friction:

$$\nabla T = A \left(\frac{q}{\rho_s s T}\right)^m \frac{\rho_n}{s}. \tag{2}$$

Here $m$ is an exponent which Gorter and Mellink set equal to three but which we shall leave adjustable, $\rho_n$ is the normalfluid density, and $s$ is the entropy per unit mass. All parameters in Eq. 2 except the amplitude $A$ and $\rho_s$ have nearly constant finite values at $T_\lambda$. From prior experiments [12] it is known that $A$ is singular at $T_\lambda$, and we write it as $A = A_0 t^{-\alpha}$ (note that the exponent $\alpha$ used here is not to be confused with the exponent $\alpha$ of the specific heat which does not occur in the present paper). Using the known critical behavior $\rho_s \propto t^\nu$ we have

$$\rho \equiv -\frac{\nabla T}{q} \sim q^{m-1} t^{-(m\nu + \alpha)}. \tag{3}$$

We write this as

$$\rho = \left(\frac{t}{t_0}\right)^{-(m\nu + \alpha)} \tag{4}$$

with

$$t_0 = \left(\frac{q}{q_0}\right)^x. \tag{5}$$

The exponents in these equations are related by

$$m = 1 + x(m\nu + \alpha). \tag{6}$$

When the heat current was turned on, there was a small offset of about $10^{-3}$ K cm$^2$ / W due to thermal crosstalk between the thermometer temperatures even deep in the superfluid phase where the resitivity of the fluid was virtually zero. This phenomenon can be understood in terms of the two-dimensional temperature field in the sidewall close to the top and bottom plate. We took it into account by allowing a constant additive term in the fit to the data. In addition, a small relative drift between the sideplane thermometers occurred in some of the experimental runs. In order to model this effect, we initially also allowed a term linear in $t$ in the model equation. However, it turned out that this term was not needed to improve the fit and that its inclusion did not significantly influence the values obtained for the other parameters. Thus the function

$$\rho = (t/t_0)^{-(m\nu+\alpha)} + \rho_0 \tag{7}$$

was fitted to the data to obtain the parameters $m\nu + \alpha$, $t_0$, and $\rho_0$ for each $q$. In Fig. 5 the background term $\rho_0$ was already subtracted.

In Fig. 6a we show $m\nu + \alpha$ as a function of $q$. At a given $P$ the data reveal no systematic $q$ dependence. Thus for each $P$ we computed a weighted average value of $m\nu + \alpha$, and show these results together with their standard errors in Fig. 6b. There seems to be a slight tendency for $m\nu + \alpha$ to decrease with increasing pressure, and a fit of a straight line to the data yields $m\nu + \alpha = A + BP$ with $A = 2.78 \pm 0.07$ and $B = -0.006 \pm 0.005$ bars$^{-1}$. Although one might expect exponents to be universal and thus independent of pressure, this universality would prevail only asymptotically, i.e. in the limit as $t$ vanishes. It is well known that the dynamics of the superfluid transition is influenced by non-universal contributions which decay exceptionally slowly as $t$ vanishes, and thus a slightly pressure dependent *effective* exponent comes as no surprise. However, in this case the pressure dependence is not much larger than the experimental resolution, and a constant, pressure-independent exponent

$$m\nu + \alpha = 2.74 \pm 0.07, \tag{8}$$

describes the data almost within their experimental uncertainty. Thus, to simplify further analysis we adopt this value.

Equation 7 was fitted to the data once more, this time holding the exponent constant at 2.74. The results for $t_0$ are shown in Fig. 7 as a function of $q$. One sees that the data at each pressure are consistent with a powerlaw description, and that there is a mild increase of $t_0$ with pressure. A fit of Eq. 5 to these data yielded values for the exponent $x$ given in Figs. 8a. The statistical errors of $x$ are relatively large at the larger pressures. This was expected because of the temerature noise induced by the pressure regulation which was absent at SVP. The results are consistent with a pressure independent

$$x = 0.89 \pm 0.01 \ . \tag{9}$$

In the fit descrived above the pressure dependence of $\rho$ is absorbed entirely in $q_0$. The results for $q_0$, obtained by holding $x$ fixed at its best value 0.89, are shown in Fig. 8b. They have quite small statistical errors, but they seem to be subject to larger systematic errors. The origin of this problem is not clear and deserves further study. A good representation of the present data is given by the straight line in the figure, which corresponds to

$$q_0 = q_{0,0} + q_{0,1} P \tag{10}$$

with $q_{0,0} = 401$ W cm$^{-2}$ and $q_{0,1} = -5.0$ W cm$^{-2}$ bar$^{-1}$.

## V. SUMMARY AND DISCUSSION

We measured the thermal resistivity $\rho$ of He$^4$ below $T_c(q)$ at several pressures between SVP and the melting curve in the heat-flux range $3 < q < 70 \mu$W/cm$^2$ and the reduced-temperature range $1 - T_c(q)/T_\lambda < t < 2 \times 10^{-5}$. As was observed before at SVP [1], we found that $\rho$ has an incipient singularity at $T_\lambda = T_c(q=0)$, and that it remains finite at $T_c(q) < T_\lambda$ where the transition to a highly dissipative state intervenes. At constant $t$ it was found that $\rho$ increases by about a factor of 2.6 with increasing pressure over the range from SVP to 29 bar. The singularity at $T_\lambda$ could be described by the powerlaw $\rho = (t/t_0)^{m\nu+\alpha}$ with $t_0 = (q/q_0)^x$. The data suggest a slight pressure dependence for the exponent $m\nu + \alpha$, but are not really good enough to establish this dependence with certainty. A pressure dependent *effective* exponent would not be surprising for a transport property near $T_\lambda$ and has been observed before, for instance, for the thermal conductivity above $T_\lambda$ [13]. Within our experimental resolution a good representation of all the data is obtained with pressure independent exponents $m\nu + \alpha = 2.74$ and $x = 0.89$ and a pressure dependent amplitude $q_0 = q_{0,0} + q_{0,1}P$ with $q_{0,0} = 401$ W cm$^{-2}$ and $q_{0,1} = -5.0$ W cm$^{-2}$ bar$^{-1}$. The corresponding value of the Gorter-Mellink exponent $m$ is 3.44. This value is slightly smaller than the previously reported result at SVP [1] $m = 3.53 \pm 0.02$, but the difference is still within the combined uncertainties of the two sets of data. In any case, both results differ significantly from the value $m = 3$ suggested by Gorter and Mellink [2] and by Vinen [3] and found in measurements at much lower temperatures.

Values of $m$ exceeding 3 are not new in the experimental literature, dating back at least three decades. [12,14] More recent work, mostly well below $T_\lambda$, is reviewed by Tough [15] and by Donnelly. [16] As an explanation, one might consider that heat transport near the superfluid transition is influenced by fluctuation effects and that the exponents are renormalized. However, recently Haussmann examined the influence of fluctuations on mutual friction using renormalization–group–theoretical methods [17]. As discussed in more detail before [1], his work predicts that the value of $m$ is unaltered by the fluctuations and his predictions do not agree very well with the experiments.

A classical model for mutual friction based on dimensional analysis of the equations of motion of vortex lines and on scaling arguments was proposed by Swanson and Donnelly (SW). [18] It yielded values of $m$ that diverge, or at least increase dramatically, near $T_\lambda$. The predicted significant dependence of $m$ on the reduced temperature $t$ seems at variance with the value $m \simeq 3.5$ found over the large range $10^{-6} \lesssim t \lesssim 10^{-3}$ and $10^{-5} \lesssim q \lesssim 10^{-2}$ W/cm$^2$ in the combination of the present work (see also Ref. [1]) and the measurements of Leiderer and Pobell [14]. However, a quantitative comparison with the SW model has not yet been carried out.

## VI. ACKNOWLEDGMENT

This work was supported by NASA through Grant No. NAG8-1757.

**Figure Captions**

Fig. 1. Schematic drawing of the cryostat. Thermal links are represented by dashed lines and liquid $^4$He supply lines by heavy solid lines.

Fig. 2. Schematic diagram of the sample cells. Shown are the mounting post (A), cell top (B), wings (C) and (G) and sideplanes (D) for mounting thermometry, stainless steel sidewall (E) and cell bottom (F).

Fig. 3. Time series during pressure regulation at (a) SVP and (b) 28.8 bars, the corresponding probability distributions (c), and the corresponding power spectral densities (d). In (c) the solid lines are fits of a Gaussian function to the data, and the wide (narrow) data set is for SVP (28.8 bars). In (d) the solid (dashed) curve is for SVP (28.8 bars).

Fig. 4. (a): A complete experimental run at a pressure of 28.8 bars. The solid, dotted, and dash-dotted traces represent the temperatures of the cell top, cell side top, and cell side bottom respectively. See text for description of letter labels. (b): An expanded view of a different high-$q$ ramp. The vertical line represents the time at which the cell bottom reached $T_c$.

Fig. 5. Results for the resistivity $\rho$. (a) is for saturated vapor pressure, and from left to right the data sets are for $q = $ 15.8, 23.7, 31.6, 39.5, 47.4, 55.2, 63.1, and 71.0 $\mu$W/cm$^2$. (b) is for 23.7 $\mu$W/cm$^2$, and from left to right the data sets are for 0.05 (SVP), 7.96, 14.7, and 28.8 bar.

Fig. 6. Results for the exponent $m\nu + \alpha$ (a) as a function of the heat-current density $q$ and (b) as a function of the pressure $P$. Open circles: saturated vapor pressure. Solid circles: 7.96 bars. Open squares: 10.8 bars. Solid squares: 14.7 bars. Open triangles: 21.6 bars. Solid triangles: 28.8 bars. In (b) the average values for all $q$ at a given pressure are shown.

Fig. 7. Results for the parameter $t_0$ in Eq. ?? obtained from a fit to the data for $\rho$ with $m\nu + \alpha$ fixed at 2.74. Open circles: SVP. Solid circles: 7.96 bars. Open squares: 10.8 bars. Solid squares: 14.7 bars. Open triangles: 21.6 bars. Solid triangles: 28.8 bars.

Fig. 8. (a): Results for the exponent $x$ obtained form a fit of Eq. 5 to the data for $t_0$ shown in Fig. 7. (b): Results for $q_0$ obtained from a fit of Eq. 5 to the data with $x$ fixed at its mean value 0.89.

FIGURES

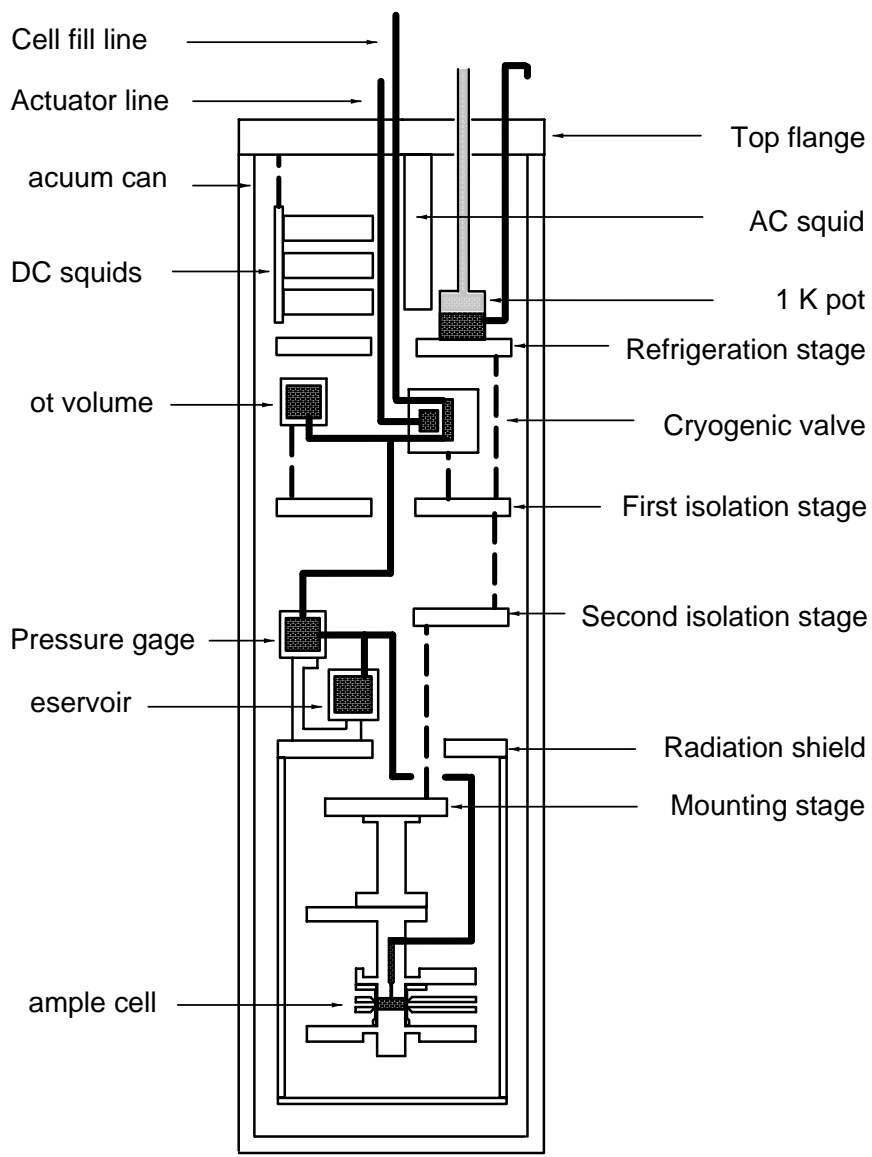

FIG. 1.

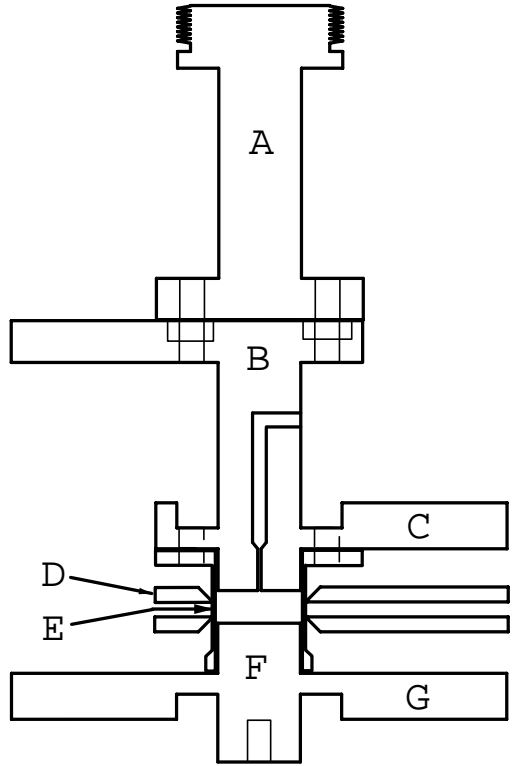

FIG. 2.

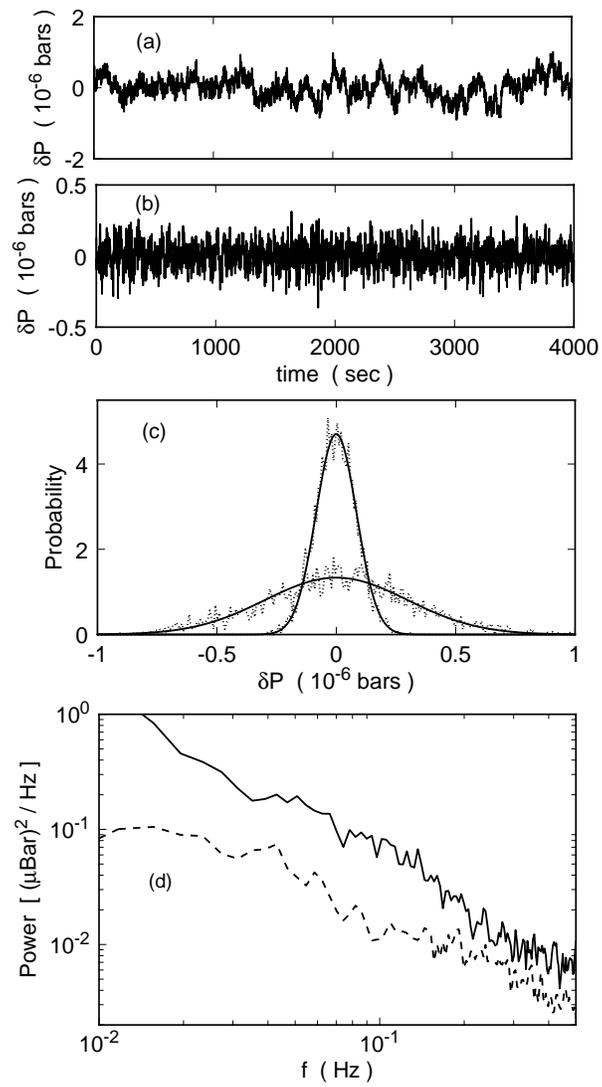

FIG. 3.

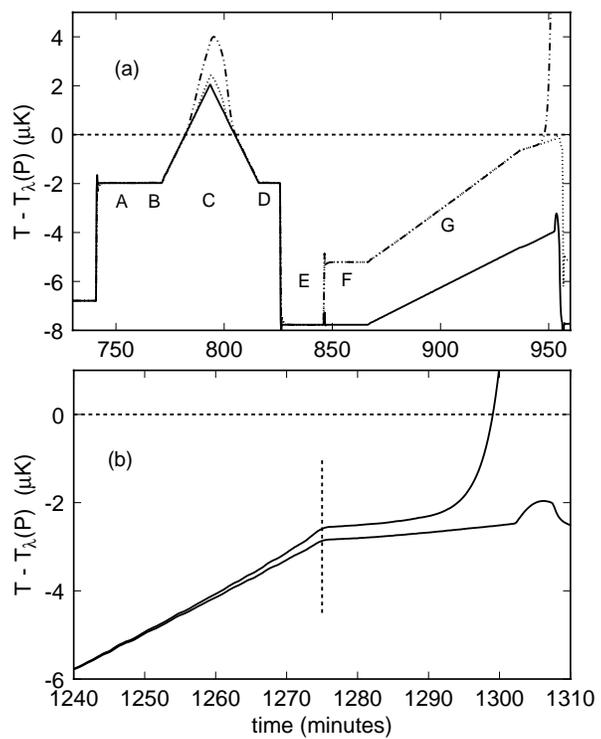

FIG. 4.

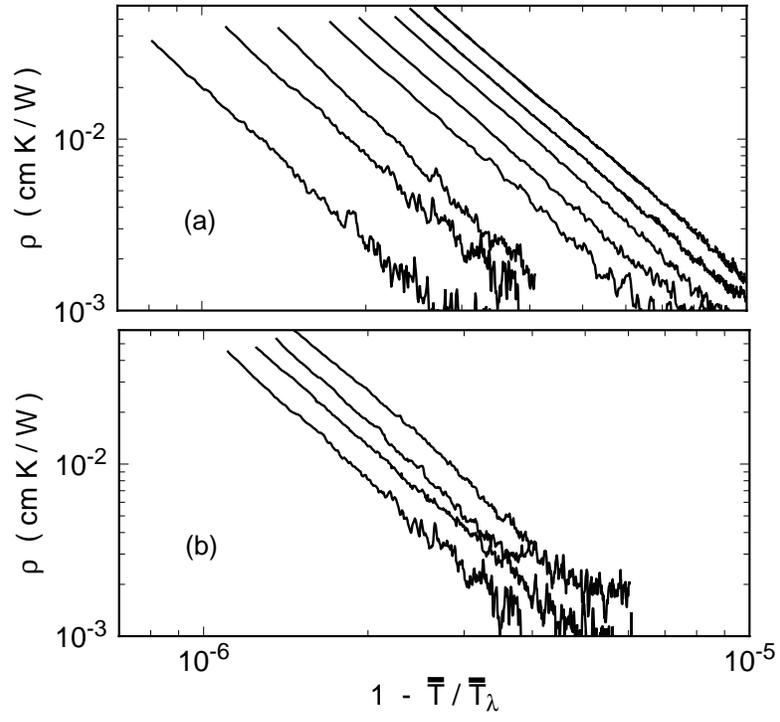

FIG. 5.

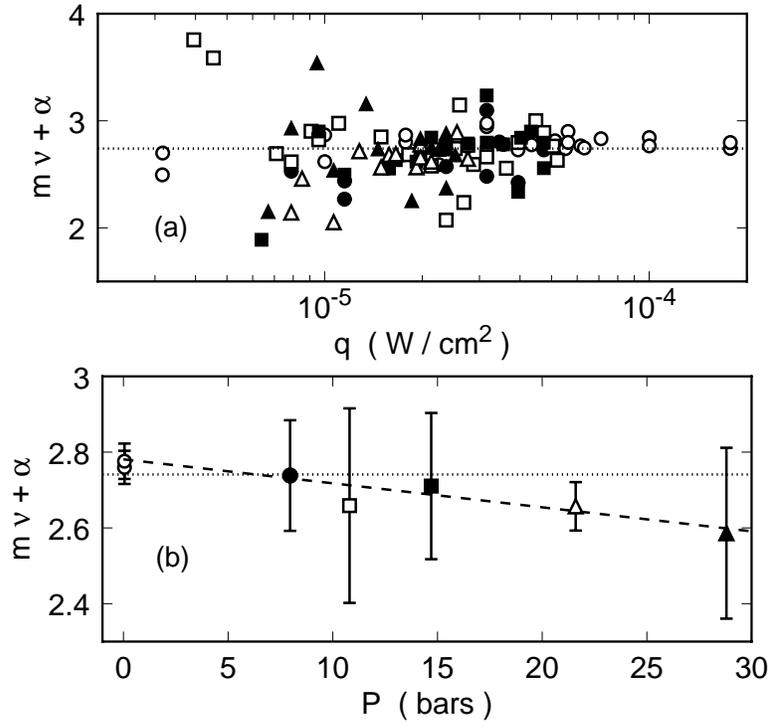

FIG. 6.

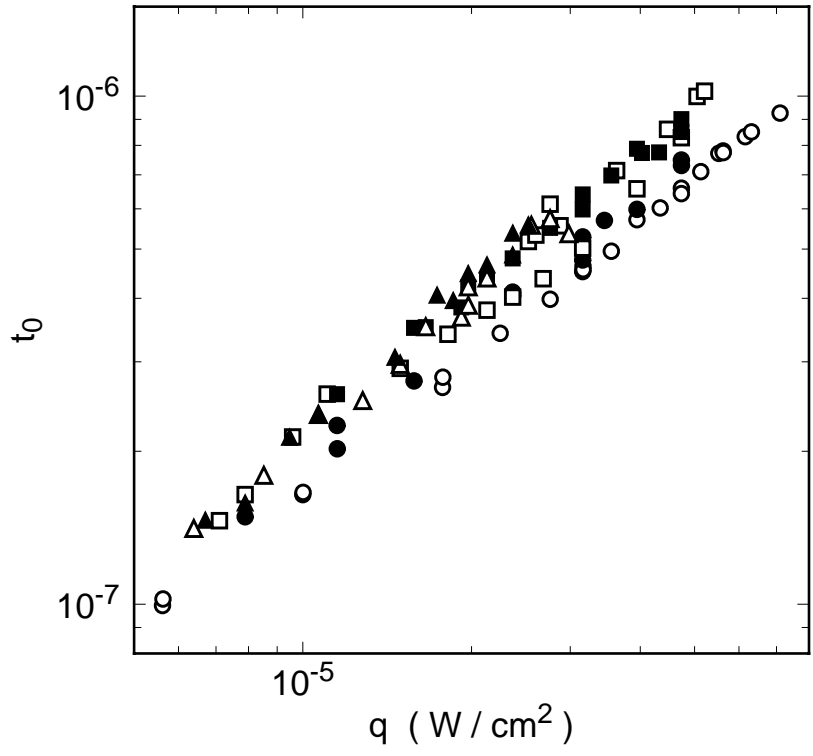

FIG. 7.

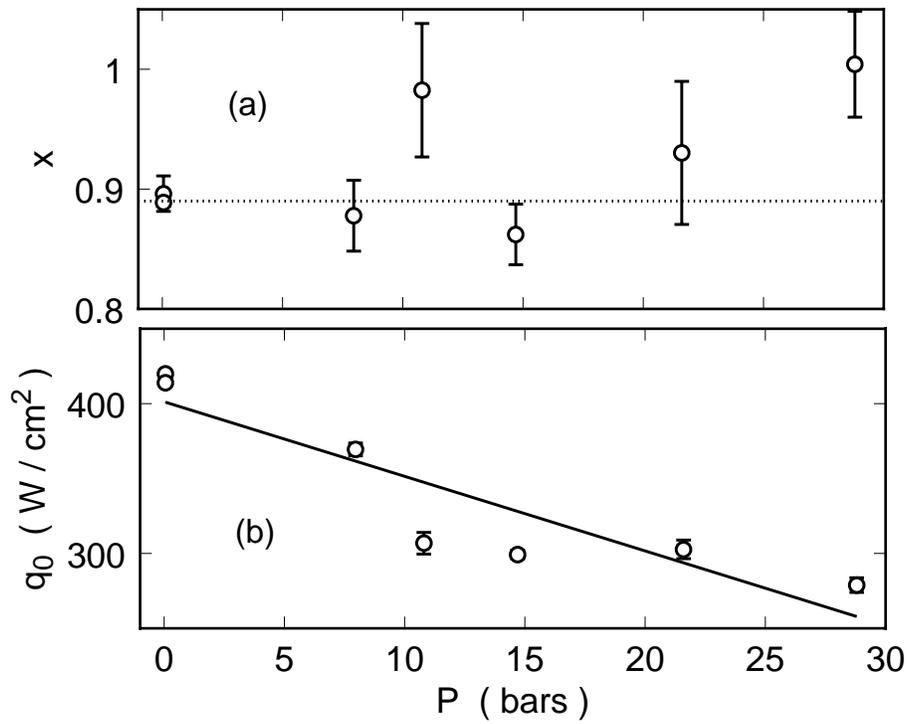

FIG. 8.